\documentclass[aps,preprint,amsfonts,amsmath,endfloats]{revtex4}
\usepackage{graphicx}  

\begin{document}
\author{Martin J.~Greenall}
\affiliation{Institut Charles
  Sadron, 23, rue du Loess, 67034 Strasbourg, France}
\author{Carlos M.~Marques}
\affiliation{Institut Charles
  Sadron, 23, rue du Loess, 67034 Strasbourg, France}
\title{Hydrophobic droplets in amphiphilic bilayers: a coarse-grained mean-field theory study}
\begin{abstract}
Hydrophobic molecules such as oils and certain drugs can be
encapsulated between the two leaflets of an
amphiphilic bilayer in both lipid and polymer systems. We
investigate the case where the hydrophobic molecules are
incompatible with the amphiphile tails and so form
droplets. Using a coarse-grained mean-field model (self-consistent
field theory, or SCFT), we find that droplets of a wide range of
sizes have the same characteristic lens shape, and explain this result
in terms of simple capillarity arguments, consistent with the measured
variations of surface concentrations of amphiphile in the bilayer and
in the monolayers that cover the droplet. We study the effect of the strength $\chi_\text{BO}$ of the repulsion
between the hydrophobic liquid and the amphiphile tails on the droplet
shape, and find a gradual flattening of the droplet as $\chi_\text{BO}$ is
reduced. The droplet remains at least metastable even at very low
values of $\chi_\text{BO}$. This is in contrast to the behavior as the length of
the hydrophobic molecules is varied. Specifically, if these molecules are at least as long as
the amphiphile tails, increasing their length further is found to have
little effect on the droplet shape, while reducing their length below
this value quickly causes the droplet to become unstable.
\end{abstract}
\maketitle

\section{Introduction}

Amphiphilic molecules in solution form
bilayers for a wide range of molecular architectures and experimental
conditions \cite{zasadzinski_rev}. These structures may form both from lipids
\cite{zidovska,sorre} and from block copolymers
\cite{jain_bates,discher,discher_eisenberg}, and are of great importance in a number of
scientific disciplines. For example, lipid bilayers are an integral
component of cells, where they form the outer membrane and also play a
role in transport processes \cite{sorre}. The bilayer vesicles that
form from block copolymers, on the other hand, are longer-lived and less permeable than
their lipid counterparts \cite{discher} making them promising
candidates as vehicles for drug delivery \cite{smart}.

The encapsulation of hydrophobic
molecules between the two leaflets of the bilayer has been a recurrent
issue, and has been discussed in a variety of contexts, including drug carrier design, vesicle
formation from inverse phase methods, and lipid distribution in cell membranes. In particular, it
is often important to know whether the hydrophobic molecules form
droplets or are spread more evenly throughout the bilayer center. For
example, lipid cell membranes are often found to contain a second
species of lipid \cite{hamilton,hamilton2,hamilton3}, in surprisingly
high concentrations \cite{khandelia}. The question then arises of whether this
is due to the presence of small lipid domains within the
bilayer, the existence of which has also been suggested to explain
the formation of the lipid droplets seen in the center of cells
\cite{ohsaki,zanghellini}. Observations of such bilayer domains have indeed
been reported in the
biophysics literature \cite{ferretti,may}, and these structures have recently been
seen in molecular dynamics simulations \cite{khandelia}. However, their probable
small size and the existence of other lipid domains in the cell
complicates the interpretation of experiments, and the issue is not
fully resolved \cite{hakumaki}.

The formation of oil droplets in amphiphilic bilayers is also a
problem of current interest in microfluidics, and has been
observed in recent experiments on block copolymer systems
\cite{hayward}. The aim of this research
\cite{hayward,shum,thiele} was to produce aqueous solutions of
monodisperse vesicles from block copolymers in water-oil-water double emulsions by
evaporating the oil. Here, the presence of oil droplets in the
bilayers is undesirable, as it leads to unevenness in the vesicle
wall. 

In addition, the encapsulation of hydrophobic molecules in bilayers is
of importance in the delivery of drugs using block copolymer vesicles
\cite{onaca,meng,li,mueller}. Although it might at first
appear more natural to encapsulate a hydrophobic substance in the core
of a spherical micelle, copolymer vesicles can offer certain
advantages over these smaller structures. In particular, they can
encapsulate both hydrophobic and hydrophilic compounds
\cite{qin,zasadzinski}. Furthermore, faster
release of the hydrophobic compound can be obtained from
vesicles \cite{chen}.

In this paper, we investigate how much information about hydrophobic droplets in bilayers can be
obtained from a simple mean-field model of oil and amphiphile in
solution. First, we will study the shape of the
droplet and to what extent this varies with its size. We will then relate
our results to capillarity arguments. Next,
we will investigate the effect of the strength $\chi_\text{BO}$ of the repulsion
between the hydrophobic liquid and the amphiphile tails on the droplet
shape. This question is of relevance to several of the situations described above, and our results will give some guidance as to how
robust the phenomenon of droplet formation is expected to be in experiments. In addition, understanding the
role of the interaction strength might allow an oil to be chosen to encourage or
discourage \cite{hayward} the formation of well-defined droplets. Finally, with similar objectives in mind, we will study
the effect of the length of the hydrophobic molecules on the droplet shape and stability.

The paper is organized as follows. In the following section, we
introduce the theoretical technique to be used, self-consistent field
theory. We then present and discuss our theoretical results,
and give our conclusions in the final section.

\section{Self-consistent field theory}\label{scft}

Self-consistent field theory (SCFT) \cite{edwards} has been used
with success to investigate the equilibrium structures formed in
melts and blends of polymers \cite{maniadis,drolet_fredrickson,matsen_book}, and may also be used to study
metastable structures, \cite{duque,katsov1} and amphiphiles in solution
\cite{cavallo,schuetz}. It can be applied to a wide range of
amphiphilic molecules, including
simple homopolymers \cite{werner}, more complex copolymers
\cite{mueller_gompper,wang} and any given mixture of these \cite{denesyuk}. SCFT requires less computational power than simulation
methods such as Monte Carlo, yet often provides comparably accurate
predictions of the form of individual structures
\cite{cavallo,wijmans_linse,leermakers_scheutjens-shape}. Furthermore, as a
coarse-grained model, with a simple description of the polymer molecules, it will allow us to capture the
basic phenomenology of the system clearly.

We now give a brief introduction to SCFT, and refer the reader to reviews
\cite{matsen_book,fredrickson_book,schmid_scf_rev} for a fuller
presentation. A complete description of our calculations for
amphiphiles in solution is given in a
recent publication \cite{gg}, and we present details only when our
current system differs from that described there. SCFT models individual molecules as random walks
in space, and so neglects fine details of their structure and packing \cite{schmid_scf_rev}. An ensemble
of many such molecules is considered. The interactions between the molecules are
modeled by assuming that the blend is incompressible and by introducing
contact potentials between the molecules \cite{matsen_book}. The
strengths of the potentials between the various species are specified by
the Flory parameters $\chi_{ij}$ \cite{jones_book}. The computational difficulty of the problem is then
sharply reduced by making a mean-field approximation
\cite{matsen_book}; that is, by neglecting fluctuations. This
approximation is quantitatively accurate when the molecules are long
\cite{cavallo,fredrickson_book,matsen_book}. In addition, SCFT can
provide considerable qualitative insight into systems containing smaller
molecules, such as lipid bilayers \cite{katsov1,gg} and aqueous
solutions of copolymer \cite{schuetz,schuetz2}.

We
now discuss the application of SCFT to our system of amphiphile and
oil in a solvent, which we model by a
mixture of block copolymer with two incompatible homopolymers that
represent the oil and the solvent respectively. Although such a
mixture of polymers may appear quite simple, models of this level of complexity
have been used to study a wide range of lipid and copolymer systems
\cite{katsov1,gbm_jcp}, and can capture broad phenomenology more clearly than
more complicated theories. We take the copolymer to have a mean-squared
end-to-end distance of $a^2N$, where $a$ is the monomer length and $N$
is the degree of polymerization \cite{matsen_book}. One half of the
monomers in this polymer are hydrophilic (type A) and the other half are hydrophobic
(type B), so that the degrees of polymerization $N_\text{A}$ and $N_\text{B}$
for the A and B blocks are
equal. We choose the same value of $a^2N$ for the A
homopolymer solvent as for the copolymers. Together with the values of
$N_\text{A}$ and $N_\text{B}$, this ensures that the amphiphile
preferentially forms flat bilayer structures \cite{gg} for the
interaction strength we will consider here. The degree of
polymerization $N_\text{O}\equiv \alpha N$ of the oil will be varied between
$N/4$ and $2N$.

In this paper, we keep the amounts of copolymer and homopolymer in the
simulation box fixed; that is, we work in the canonical ensemble. This
will make it easier for us to access more complex
structures such as droplets. Such structures are
more difficult to stabilize in ensembles where
the system is able to relax by
varying the amount of the various species, and can require constraints
to be imposed on the density profile
\cite{katsov1}.

For concreteness and to introduce the appropriate notation, we note
that the SCFT approximation to the free energy of our system has the form
\begin{align}
\lefteqn{\frac{FN}{k_\text{B}T\rho_0V}=\frac{F_\text{h}N}{k_\text{B}T\rho_0V}}\nonumber\\
& 
-(1/V)\int\mathrm{d}\mathbf{r}\,[\chi N_\text{AB} (\phi_\text{A}(\mathbf{r})+\phi_\text{S}(\mathbf{r})-\overline{\phi}_\text{A}-\overline{\phi}_\text{S})(\phi_\text{B}(\mathbf{r})-\overline{\phi}_\text{B})
\nonumber\\
&
+\chi
N_\text{AO}(\phi_\text{A}(\mathbf{r})+\phi_\text{S}(\mathbf{r})-\overline{\phi}_\text{A}-\overline{\phi}_\text{S})(\phi_\text{O}(\mathbf{r})-\overline{\phi}_\text{O})
\nonumber\\
&
+\chi N_\text{BO}(\phi_\text{B}(\mathbf{r})-\overline{\phi}_\text{B})(\phi_\text{O}(\mathbf{r})-\overline{\phi}_\text{O})]
\nonumber\\
& 
-(\overline{\phi}_\text{A}+\overline{\phi}_\text{B})\ln (Q_\text{AB}/V)-\overline{\phi}_\text{S}\ln(Q_\text{S}/V) -(\overline{\phi}_\text{O}/\alpha)\ln (Q_\text{O}/V)
\label{FE}
\end{align}
where the $\overline{\phi}_i$ are the mean volume fractions of the
various components. The $\phi_i(\mathbf{r})$
are the local volume fractions, with $i=\text{A}$ for the hydrophilic blocks,
$i=\text{B}$ for the hydrophobic blocks, $i=\text{O}$ for the oil, and
$i=\text{S}$ for the solvent. The first Flory parameter, $\chi_\text{AB}$, is set to $50/N$,
so that sharp, well-defined bilayers form. The other Flory parameters will be
varied to study the effect of the nature of the oil on the droplet
shape. $V$ is the
total volume of the system, $1/\rho_0$ is the volume of a monomer, and $F_\text{h}$
is the SCFT free energy of a homogeneous system containing the same
components. The details of the individual polymers enter through the
single-chain partition functions $Q_i$. These are calculated
\cite{matsen_book} from
integrals over the
propagators $q$ and $q^\dagger$, which are also
used to compute the polymer density
profiles \cite{matsen_book,fredrickson_book}. Reflecting the
fact that the molecules are modeled as random walks, the propagators
satisfy modified diffusion equations with a field term that describes
the polymer interactions. These equations are solved using a finite difference method
\cite{num_rec} with step size of $0.04\,aN^{1/2}$. We assume that the
droplet is cylindrically symmetric and forms at the center of the system, and hence
consider an effectively two-dimensional problem in a cylindrical
calculation box. Reflecting boundary
conditions are imposed at the edges of the system.

The derivation of the mean-field free energy $F$ also generates a set of simultaneous equations linking
the values of the fields and densities. In order to calculate the SCFT density profiles for a given set of
polymer concentrations, we begin by making an initial guess
for the fields $w_i(\mathbf{r})$ and solve the diffusion
equations to calculate the propagators and then the densities
corresponding to these fields. The new $\phi_i(\mathbf{r})$ are then
substituted into the simultaneous equations to calculate new values
for the $w_i$ \cite{matsen2004}. The procedure is repeated until
convergence is achieved. We have checked that the algorithm converges
to the same solution from different initial states and with different
iteration speeds.

To form the structure we wish to study, a bilayer in the $z=0$ plane encapsulating a droplet at its center, a
suitable initial guess for the $w_i$ must be made. It is important to note that it is
not necessary to include any detailed information about the shape of
the droplet or bilayer in this ansatz. Although, for the sake of speed, we will
often use the final self-consistent fields corresponding to one
bilayer-droplet system as initial guesses for a subsequent
calculation, the initial form of the fields can be very
simple. Specifically, to encourage the formation of the
bilayer, it is sufficient to start the SCFT iteration with a simple
square well for the hydrophobic block field $w_\text{B}$. If $z_0$ is the approximate width of the bilayer, we set $w_\text{B}$ to a low value for
$-z_0/2<z<z_0/2$, and a higher value elsewhere. The initial value of
the hydrophilic block
field $w_\text{A}$ can simply be set to zero, as the A and B blocks
are connected and the above ansatz for $w_\text{B}$ is enough for
an AB bilayer to form. The iteration for the field corresponding to the
oil, $w_\text{0}$, can similarly be initiated with a square well
potential. The
potential is set to a low value in a cylindrical region at the center
of the system
($-z_0'<z<z_0'$, $r<r_0$), with $z_0'<z_0$ and $r_0$ much smaller
than the radius of the simulation box.

We set the tension of a bilayer with no oil to zero, as
this corresponds most closely to the experimental situation of a
vesicle in solution. To find the zero-tension bilayer, we proceed as follows \cite{gbm_jcp}. First, we
calculate the free-energy
density of a (one-dimensional) box containing an infinite planar aggregate
in solvent.
The volume of the simulation box is then varied in the direction
perpendicular to the bilayer surface, keeping the volume fraction of copolymer
fixed (at $10\%$), until the box size with the minimum free-energy
density is found. This scheme was introduced to mimic the
behavior of a system of many aggregates \cite{gbm_jcp,gbm_macro}, which minimizes its
free energy by varying the number of aggregates and hence the volume
(`box size') occupied by each. Equivalently, this procedure allows us
to prepare a bilayer under zero tension. Decreasing the
box size at constant copolymer volume fraction reduces the amount of
amphiphile in the system and so thins the bilayer,
which corresponds to stretching it parallel to its
surface. Conversely, increasing
the box size thickens the bilayer, which is physically equivalent to
compressing it. The bilayer found for the box size where the free
energy is at a minimum is one that is neither too stretched nor too
compressed and has no contributions to the free energy from
polymer chains that are forced into unfavorable configurations.

This calculation is then used to fix the size of the cylindrical box
in the $z$-direction to $17.6\,aN^{1/2}$, so that $-Z<z<Z$, where $Z=8.8$. The radius of the box is set to $R=16\,aN^{1/2}$, to allow droplets of a wide
range of sizes to be studied.

\section{Results and discussion}\label{results}
In this section, we study the shape and size of the droplets in detail for a single set
of system parameters. Then we relate the droplet shape to the
bilayer and monolayer tensions and amphiphile concentrations. Finally,
we investigate how the droplet shape and stability depends on the nature
and size of the oil molecules.
\subsection{Droplet shape.}
To begin, we calculate the density profiles for bilayer-encapsulated droplets of various
sizes. We focus on a system with oil molecules that are half the size
of the amphiphiles, so that $N_\text{O}=N/2$. The interaction strength between
the hydrophobic B-block of the amphiphile and the oil is set to
$\chi_\text{BO}=5/N$. The effect of varying this quantity will be
investigated later. Given that the
strength of the repulsion between the A and B blocks of the amphiphile
has already been set to $\chi_\text{AB}=50/N$, we no longer have
complete freedom in our choice of the final Flory parameter,
$\chi_\text{AO}$. If we assume that $\chi_{ij}$ is related to the
polarizabilities $\alpha_{i,j}$ of the two polymer species by
$\chi_{ij}=\kappa(\alpha_i-\alpha_j)^2$, where $\kappa$ is a
constant of proportionality \cite{boudenne_book}, we find that
$\chi_\text{AO}$ is given in terms of the other two
interaction strengths by
\begin{equation}
\chi_\text{AO}=\chi_\text{AB}\left(1\pm\sqrt{\frac{\chi_\text{BO}}{\chi_\text{AB}}}\right)^2
\label{flory_eqn}
\end{equation}
and is therefore set to $\chi_\text{AO}=23.4/N$, where we choose the
negative root to give a moderate incompatibility between the oil and
the solvent.

\begin{figure}
\includegraphics[angle=270,width=\linewidth]{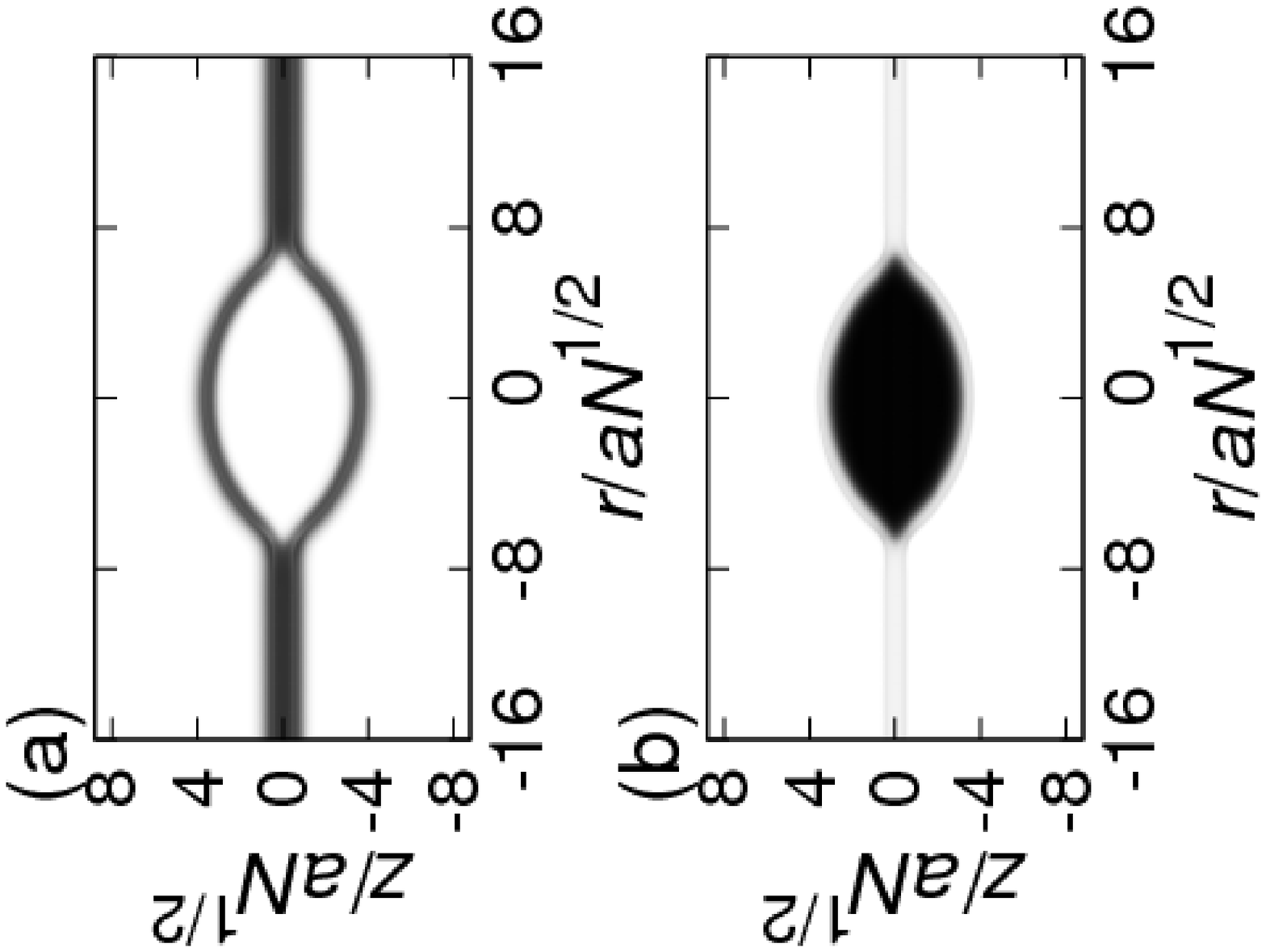}
\caption{\label{dropcomb_fig} Density plots of (a) amphiphile (including both
  hydrophilic and hydrophobic blocks) and (b) oil in a droplet-bilayer
  system with $\overline{\phi}_\text{O}=0.04$. Cylindrical polar coordinates are used, and dark regions indicate high
volume fraction.}
\end{figure}

Starting with the tensionless bilayer described above, we compute
the density profiles of bilayer-encapsulated droplets for a range of
oil volume fractions between $\overline{\phi}_\text{O}=0.01$ and
$\overline{\phi}_\text{O}=0.1$. The first of these values
corresponds to the smallest droplet that could be stabilized in our
calculations. In Figure \ref{dropcomb_fig}, we show the density profiles for
(a) the amphiphile and (b) the oil for an intermediate-sized droplet
with $\overline{\phi}_\text{O}=0.04$. Figure \ref{dropcomb_fig}a clearly
shows the splitting of the amphiphile bilayer into two thin monolayers
to incorporate the droplet. In Figure \ref{dropcomb_fig}b, we plot the
density profile of this lens-shaped droplet, and see also that a
significant amount of oil remains between the two leaflets of the
bilayer in the region
surrounding the drop. This feature appears as horizontal gray lines on
either side of the droplet
in the density plot of Figure \ref{dropcomb_fig}b, and is a result of the
relatively weak repulsion $\chi_\text{BO}$ between the hydrophobic
block and the oil. 

\begin{figure}
\begin{center}
\includegraphics[width=0.4\linewidth]{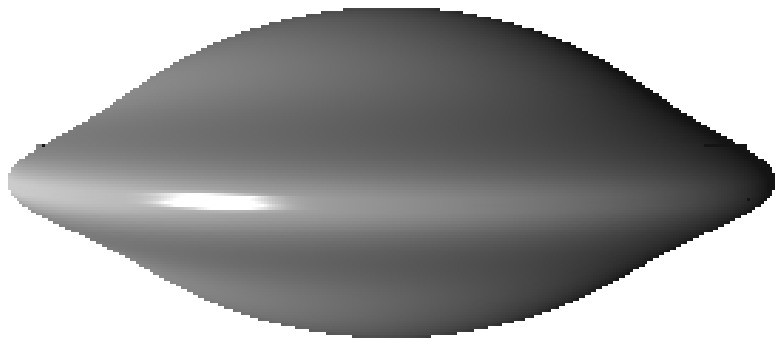}
\end{center}
\caption{\label{lathe_fit_bw_fig} Ray-traced plot of the surface of
  the droplet shown in Figure \ref{dropcomb_fig}b.}
\end{figure}

To help visualize the droplets, we show a ray-traced plot of the
droplet surface (defined as the locus of points where
$\phi_\text{O}(\mathbf{r})=0.5$) in Figure \ref{lathe_fit_bw_fig}. The
overall lens shape of the droplet is clearly visible, as is the slight
rim where the edge of the droplet meets the oil remaining in the bilayer.

\begin{figure}
\includegraphics[width=\linewidth]{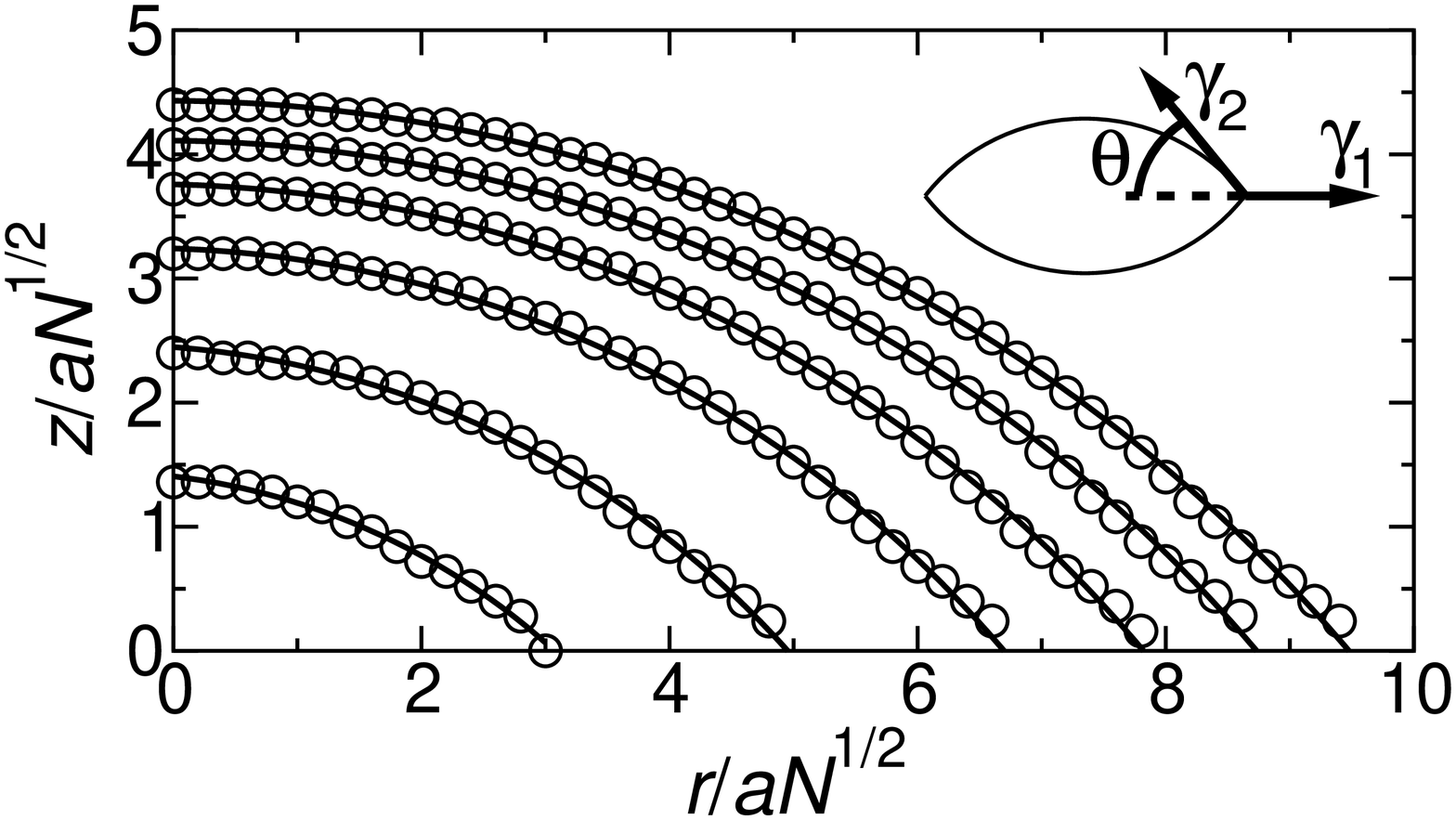}
\caption{\label{circlefits_fig} Cuts through the droplet surface
  for a range of droplet sizes with
  $\overline{\phi}_\text{O}=0.01$, $0.02$, $0.04$, $0.06$, $0.08$, and
  $0.1$. Circles show the data points from our SCFT
  calculations. Although all points are used in the fits, only every
  fifth point is plotted for clarity. Solid lines show fits to the
  data using sections of a circle. The inset shows the contact angle
  $\theta$ and the surface tensions $\gamma_1$ and $\gamma_2$, for
  later reference.}
\end{figure}

\subsection{Surface concentrations and tensions.}

In order to gain more detailed insight into the droplet shapes, we
plot cuts through the droplet surface for a range of droplet sizes from the smallest
to the largest (see Figure \ref{circlefits_fig}). If we assume that no long-range forces act in the
system, that the bending rigidity of the membrane can be neglected,
and that the pressure inside the droplet is constant, both the upper
and lower halves of the droplet will have a spherical cap shape
\cite{leger} in order to obey the Laplace law
\cite{rowlinson_widom,landau_lifshitz}. The expected
shape for these cuts shown in Figure \ref{circlefits_fig} is then a section
of a circle, with the contact angle $\theta$ (see inset) determined by
the mechanical equilibrium of the surface tensions along the contact
line \cite{leger}. We indeed find that this shape
gives a very good fit to the data for all droplet shapes
(Figure \ref{circlefits_fig}), with a slight deviation in the rim region
shown in Figure \ref{lathe_fit_bw_fig}, where the droplet spreads out
slightly between the two amphiphile leaflets instead of forming a perfect cusp. In addition, the contact angle calculated from
the fits is the same for all droplet sizes to the accuracy of the
calculations, and is given by $\theta\approx 51^\circ$.

\begin{figure}
\includegraphics[width=\linewidth]{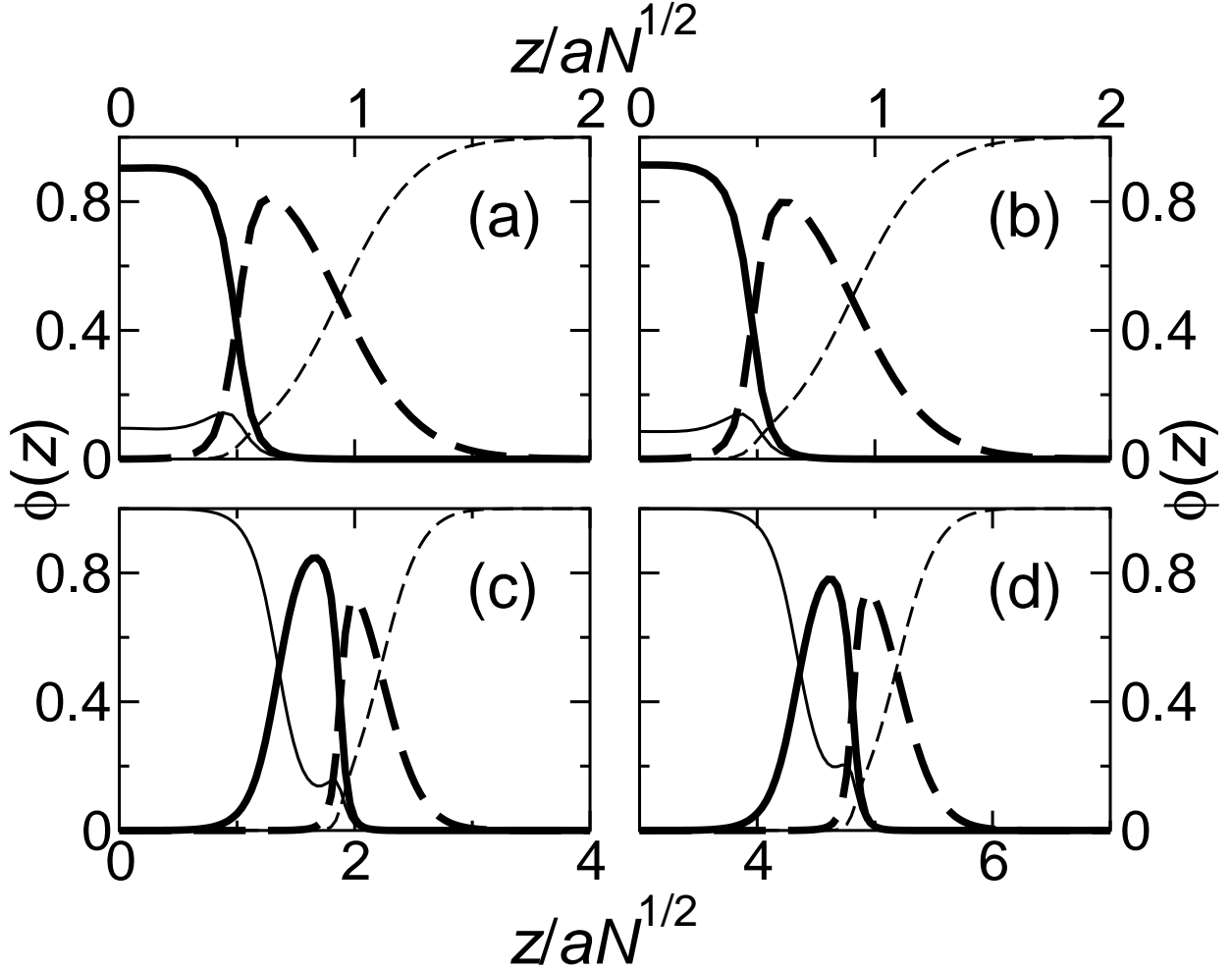}
\caption{\label{centres_edges_fig} Cuts through the density profiles
  perpendicular to the bilayer
  at the edge of the system and at the center of the droplet. The
  hydrophilic A-blocks are shown with thick dashed lines, the
  hydrophobic B-blocks with thick full lines, the oil with thin full
  lines, and the solvent with thin dashed lines. (a) Bilayer at the
  edge of a system containing the smallest droplet studied
  ($\phi_\text{O}=0.01$). (b) Bilayer at the edge of a system
  containing the largest droplet ($\phi_\text{O}=0.1$). (c) Monolayer
  covering a droplet at the center of the $\phi_\text{O}=0.01$
  system. (d) Monolayer covering a droplet a the center of the
  $\phi_\text{O}=0.1$ system. Note the change in $z$-axis between (c)
  and (d). 
  }
\end{figure}

To understand how the presence of the bilayer leads to the formation of
these lens-shaped droplets, and to illustrate some other
features of the density profiles plotted in Figure \ref{dropcomb_fig}, we now
plot a series of cuts through the density profiles of the various
species. In Figure \ref{centres_edges_fig}a, we show a cut in the
$z$-direction (perpendicular to the bilayer) through the density profiles of all species at the edge
of the system containing the smallest droplet studied. This plot shows the
state of the bilayer as far away from the droplet as possible. In
Figure \ref{centres_edges_fig}b, we show the corresponding plot for the
bilayer in the system with the largest droplet. First, we note that the two
plots are very similar and that the bilayer is not strongly distorted
by the presence of the droplet. We will return to this point in a more
quantitative fashion later on. Second, we see that a small but
significant amount of oil remains in the center of the
bilayer. Furthermore, the density profile of the oil shows a clear
peak at the interface between the hydrophilic A and hydrophobic B
blocks. This is because both $\chi_\text{AO}$ and $\chi_\text{BO}$ are
  smaller than $\chi_\text{AB}$, the repulsion between the two blocks of
  the amphiphile. A thin film of oil therefore forms in this region to
protect these two strongly incompatible species from each other.

A clear contrast is seen between the density profiles of the bilayer a long way from the droplet, and those of the
monolayer that covers the droplet. These latter profiles, calculated
at the center of the system ($r=0$), are plotted in
Figure \ref{centres_edges_fig}c (for the smallest droplet) and d (for the
largest droplet). We note that, in both these cases, the maximum
values of the density profiles of the A and B blocks of the amphiphile are lower than in
Figure \ref{centres_edges_fig}a and b. This is because the monolayer has to be
stretched and thinned to cover the droplet, lowering the surface
amphiphile concentration. In addition, the difference between the
profiles for the smallest and largest droplets is much more marked
than in the case of the bilayer. In particular, the amphiphile
concentration in the monolayer covering
the largest droplet (Figure \ref{centres_edges_fig}d) is noticeably lower than in that covering the smallest droplet
(Figure \ref{centres_edges_fig}c), showing that the monolayer must be further
stretched to encapsulate more oil. Furthermore, the monolayer in
Figure \ref{centres_edges_fig}d is more symmetric with respect to its inner
and outer leaflets than that in
Figure \ref{centres_edges_fig}c, and the peak values of the A- and B-block
concentrations are much closer. The reason for this is that the
surface of the larger droplet is flatter, and the monolayer that
encloses it is quite close to that which would form at a planar
oil-solvent interface.

\begin{figure}
\includegraphics[width=\linewidth]{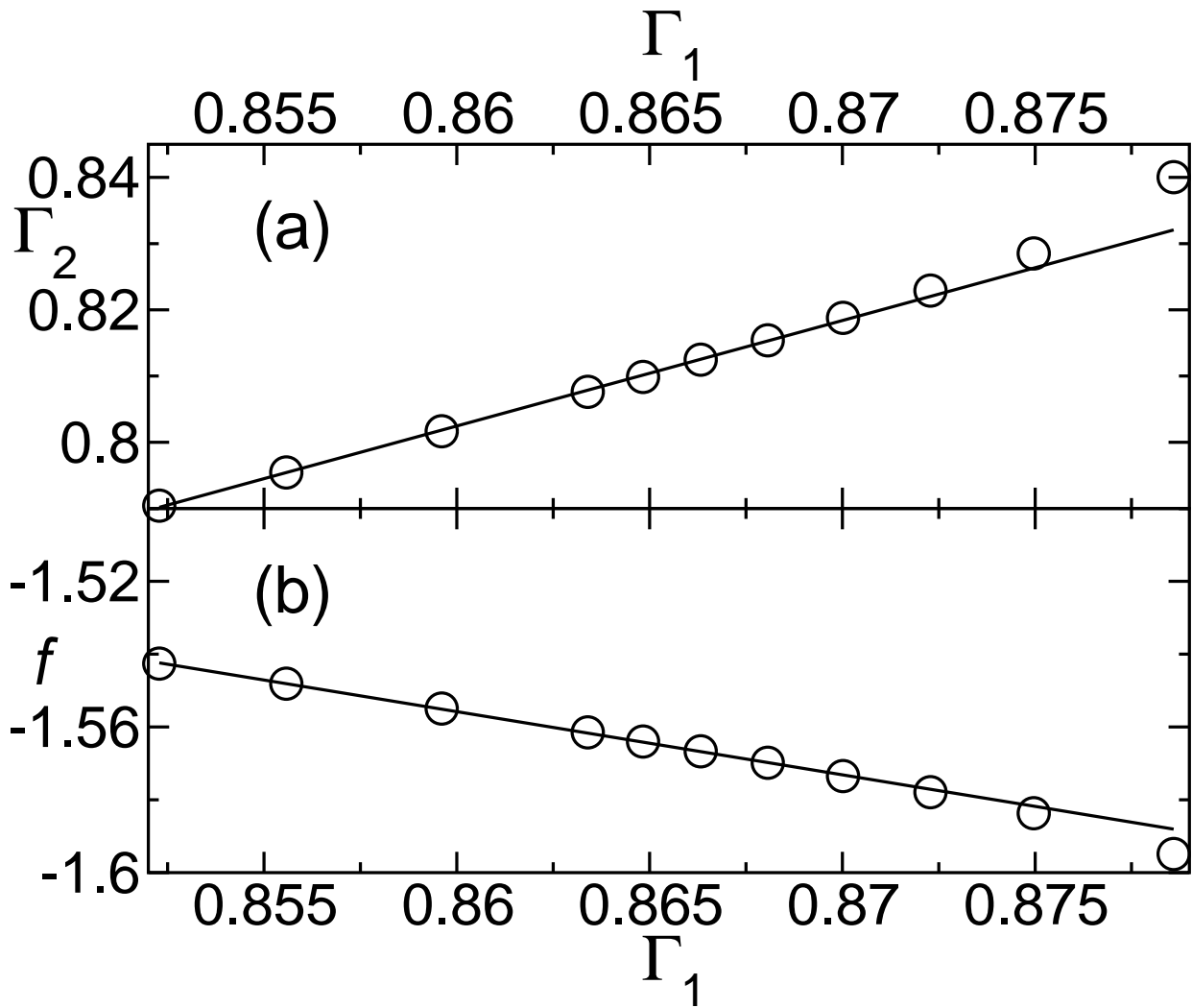}
\caption{\label{centre_vs_edge_fig} (a) Surface density of the
  monolayer covering the droplet plotted against that of (half) the
  bilayer at the edge of the system. (b) Normalized free energy
  density of the bilayer
  plotted against its surface density.
  }
\end{figure}

We now present a more quantitative discussion of the amphiphile density profiles. To begin, we integrate the bilayer and monolayer density
profiles (Figure \ref{centres_edges_fig}) in the $z$-direction, including
both hydrophilic and hydrophobic blocks. Specifically, we calculate
\begin{align}
\lefteqn{\Gamma_1=\int^Z_0\,\mathrm{d}z\left(\phi_\text{A}(R,z)+\phi_\text{B}(R,z)\right)} \nonumber\\
& 
\Gamma_2=\int^Z_0\,\mathrm{d}z\left(\phi_\text{A}(0,z)+\phi_\text{B}(0,z)\right) 
\label{surface_densities}
\end{align}
$\Gamma_1$ is then the surface density of (half) the bilayer at the
edge of the system ($r=R$), while $\Gamma_2$ is the monolayer surface
density at the center of the system ($r=0$). The amount of amphiphile
remaining in the bulk is low, so that its contribution to the surface
densities is very small. $\Gamma_1$ and $\Gamma_2$ are calculated for all
droplet sizes studied and are plotted against each other in
Figure \ref{centre_vs_edge_fig}a. As would be expected from the profiles
shown in Figure \ref{centres_edges_fig}, the monolayer surface density varies
over a wider range than the corresponding quantity for the bilayer, as
the amphiphilic molecules are spread out more and more thinly in the
monolayer as the
droplet size increases. Furthermore, $\Gamma_1$ and $\Gamma_2$  are linearly
related for a wide range of droplet sizes, with deviations from linearity only being seen for the higher surface densities
corresponding to very small droplets. To understand this
behavior, we relate $\Gamma_1$ and $\Gamma_2$ to the
corresponding surface
energies. First, we note that the amphiphile in our system acts as a
surfactant, separating the solvent from the oil in the droplet and
bilayer. Then, we assume that adding amphiphile linearly reduces the
surface tension from its value in the absence of amphiphile,
$\gamma_0$, so that $\gamma_i=\gamma_0-\delta\Gamma_i$. Here, $\delta$
is a constant of proportionality and $i=1$ for the tensions and
densities at the system edge and $2$ for those at the droplet
surface. Balancing these two tensions at the rim where the droplet
meets the bilayer, as shown in the inset to
Figure \ref{circlefits_fig}, we find also that
$\gamma_1=\gamma_2\cos\theta$. Combining this with our expressions for
$\gamma_1$ and $\gamma_2$, we have
\begin{equation}
\Gamma_2=\frac{\Gamma_1}{\cos\theta}+\frac{\gamma_0}{\delta}\left(1-\frac{1}{\cos\theta}\right)
\label{g1g2}
\end{equation}

From the slope of the straight line in Figure \ref{centre_vs_edge_fig}a, we
find that $\cos\theta\approx 0.627$, so that $\theta\approx
51^\circ$, in excellent agreement with our independent
measurement of $\theta$ from the cross-sections in Figure \ref{circlefits_fig}. 
This shows the validity of the force balance argument, and also
confirms our use of the same proportionality constant $\delta$ in our
expressions for $\gamma_1$ and $\gamma_2$. To
check our linear formula for $\gamma_1$, we have also calculated,
using Equation \ref{FE}, the free
energy density of a flat oil-containing bilayer with the profile shown
in Figure \ref{centres_edges_fig}a. Similar calculations are performed for all values of
$\overline{\phi}_\text{O}$.
We then plot the quantity
$f=FN/k_\text{B}T\rho_0V-F_\text{h}N/k_\text{B}T\rho_0V$ (the free
energy density measured with respect to that of the homogeneous
solution with the same composition) as a function of the surface
concentration $\Gamma_1$. As can be
seen from Figure \ref{centre_vs_edge_fig}b, $f$ decreases linearly with
$\Gamma_1$ for all but the very smallest droplets, confirming
our simple model for the surface energy $\gamma_1$. We note that the free energy
density as calculated from Equation \ref{FE} includes a contribution from
the solvent region as well as from the bilayer itself. However, this
is likely to have a relatively small effect on the variation of $f$, as the bulk amphiphile
concentration changes very little with droplet size.

Both plots
show some deviation from linearity for the smallest two or three droplet sizes
considered. For the plot of the two surface concentrations in
Figure \ref{centre_vs_edge_fig}a, the deviation comes from the increasing
relative importance of the rim around the edge of the droplet
(Figure \ref{lathe_fit_bw_fig}), which means that a simple
force balance argument based on a well-defined contact angle is less
valid. The slight breakdown of linearity in Figure \ref{centre_vs_edge_fig}b
may be due to the fact that the droplet is nearing its lower
size limit and the free energy density $f$ is dropping more rapidly as
the bilayer relaxes towards the flat state.

\subsection{Nature and size of the oil molecules.}

\begin{figure}
\includegraphics[width=\linewidth]{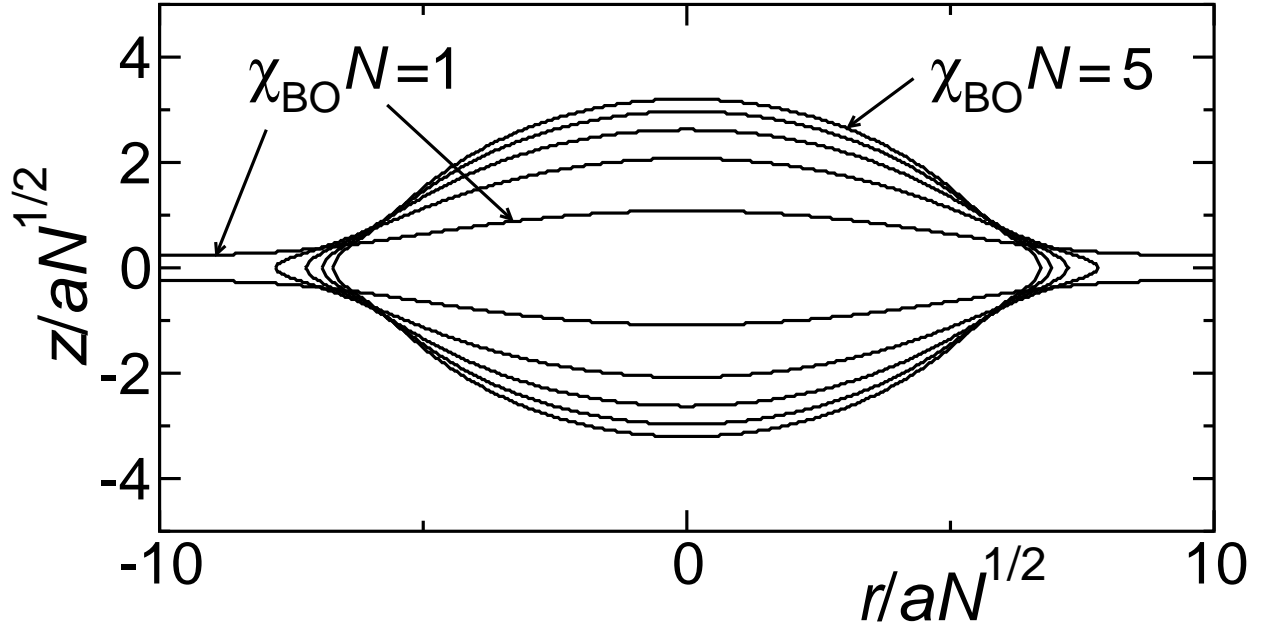}
\caption{\label{oil_chi_fig} Droplet outline for a range of values of
  $\chi_\text{BO}N$ from $1$ (flattest drop) to $5$ (roundest
  drop). The oil volume fraction $\overline{\phi}_\text{O}$ is fixed to $0.04$.
  }
\end{figure}

Having established a basic picture of droplet formation in our system,
we now turn our attention to the question of how the nature and size
of the oil molecules affects the shape and stability of the
droplets. First, we investigate the effect of changing the Flory parameter
$\chi_\text{BO}$ determining the strength of the interaction between
the oil and the hydrophobic blocks of the amphiphile. This corresponds
to changing the chemical nature of the oil while keeping the length of
the oil molecules constant. We keep the same type of amphiphiles as
used in the preceding section, so that $N$ and $\chi_\text{AB}$ are
unchanged. However, as noted earlier, the three Flory parameters
cannot be varied completely independently \cite{boudenne_book}, and $\chi_\text{AO}$ must
be recalculated according to Equation \ref{flory_eqn} for each value of
$\chi_\text{BO}$. The volume fraction of oil is set to
$\overline{\phi}_\text{O}=0.04$. In Figure \ref{oil_chi_fig}, we
plot the outlines of the droplets formed as $\chi_\text{BO}N$ is
decreased from $5$ (the value used in our earlier calculations) to
$1$ in steps of $1$. The droplet with $\chi_\text{BO}N=5$ is that with the most
rounded shape and the greatest thickness in the $z$-direction. As
$\chi_\text{BO}N$ is lowered to $4$, the droplet spreads outward
slightly into the bilayer and becomes thinner, as the oil becomes more
compatible with the hydrophobic blocks of the amphiphile. The
spreading effect here is rather small, suggesting that, above a certain
$\chi_\text{BO}N$, the droplet shape is relatively insensitive to the
nature of the oil and retains its characteristic lens form. As
$\chi_\text{BO}N$ is reduced further, the droplet continues to
spread. However, the amount by which the droplet thickness falls due
to each reduction of $\chi_\text{BO}N$ by $1$ gradually increases,
until we obtain an almost flat structure at $\chi_\text{BO}N=1$. It
is interesting to note that, even for this very weak repulsion between
the oil and the hydrophobic blocks, the droplet remains at least
metastable.

The results shown in Figure \ref{oil_chi_fig} differ somewhat
from the classical problem of the spreading of a single droplet
\cite{leger}, as our droplet is in equilibrium with a film of oil in
the bilayer, which grows in thickness, taking material from the droplet, as the oil becomes less
incompatible with the hydrophobic B-blocks. This is particularly clear for the lowest
value of $\chi_\text{BO}N$ considered. Here, the oil concentration in
the bilayer is so high that our
definition of the droplet surface as the
locus of points at which $\phi_\text{O}(\mathbf{r})=0.5$ now includes the
oil film as an extension of the droplet.

\begin{figure}
\includegraphics[width=\linewidth]{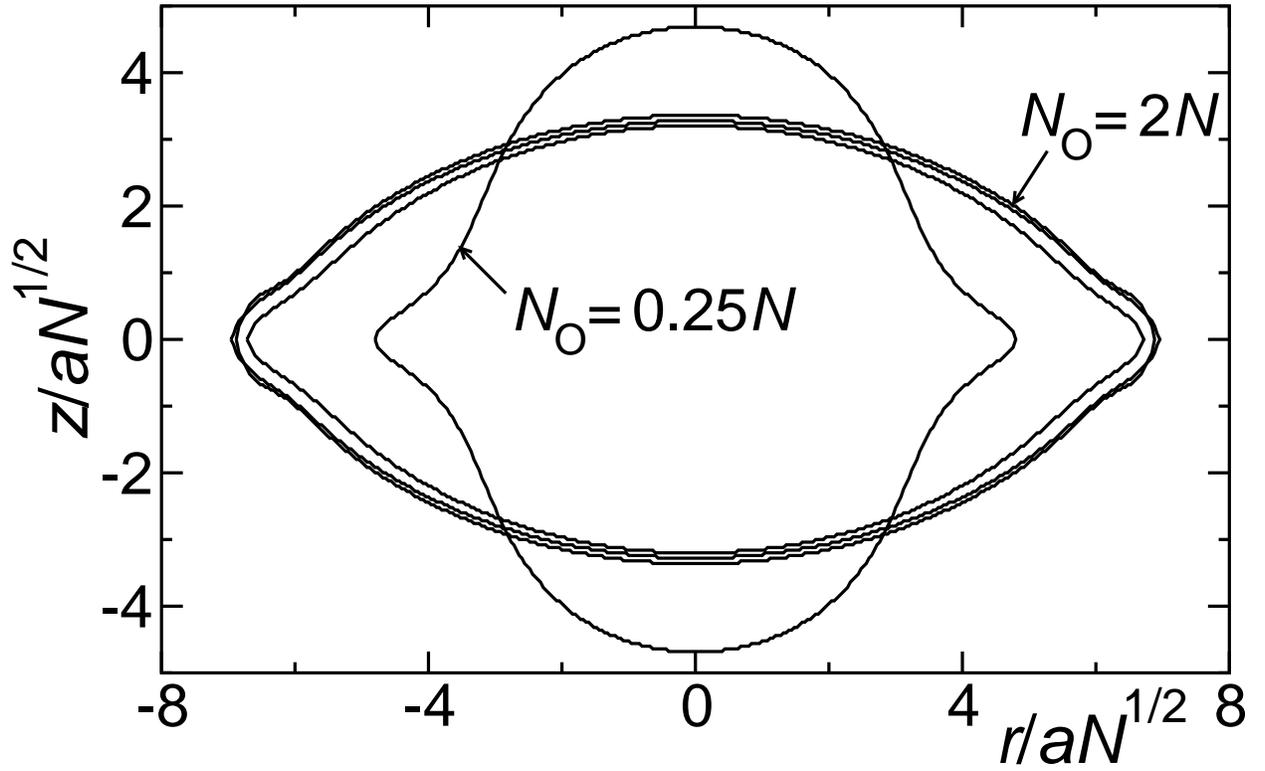}
\caption{\label{oil_size_fig} Droplet outline for a range of values of
  $N_\text{O}$ from $2N$ (outermost flat drop), through $N$ and $0.5N$, to $0.25N$ (round drop).
  }
\end{figure}

Finally, we study the effect of the size of the oil molecules on the
droplet shape. Returning to our original set of Flory parameters, with
$\chi_\text{BO}N=5$, we consider the following values of the oil
polymerization index: $N_\text{O}=2N$, $N$, $0.5N$ (the original value),
and $0.25N$, and plot the droplet outlines in Figure \ref{oil_size_fig}. As
above, $\overline{\phi}_\text{O}=0.04$. For
the largest three values of $N_\text{O}$, the droplet shape changes rather
little. It simply shrinks slightly as $N_\text{O}$ is lowered, as this
change reduces
the repulsion between the oil and the hydrophobic sections of the
amphiphile so that more material leaks out of the droplet into the
oil film.

However, as $N_\text{O}$ is lowered further, to
$0.25N$, a sharp change occurs in the droplet shape. The thickness of
the droplet in the $z$-direction is now significantly greater, while its radius
is smaller. It is difficult to interpret this result in terms of the
simple force balance arguments used earlier. This is because the rim
feature, which was previously a small perturbation on droplets whose
shape could be represented by two joined spherical caps \cite{leger},
is now a much more significant part of the droplet, since the small
oil molecules penetrate more effectively into the bilayer. We speculate that the
elongated shape of this droplet may be a precursor to its eventual
splitting into two smaller monolayer-wrapped droplets, such as those
recently studied by Kusumaatmaja and Lipowsky in the context of
membranes in contact with several fluids \cite{kusumaatmaja}. In any
case, it
certainly seems that the simple single-droplet solution to SCFT
becomes unstable around $N_\text{O}=0.25N$. If we reduce $N_\text{O}$ below
this value, no solution to the SCFT
equations can be found using our current methods, and our algorithm
slows down considerably even for $N_\text{O}=0.25N$.

\section{Conclusions}\label{conclusions}
Using a coarse-grained mean-field approach (self-consistent field theory) we have modeled
several aspects of the structure of hydrophobic droplets encapsulated between the two leaflets of an
amphilic bilayer. First, we have found that droplets of a range of
sizes have the same simple lens shape that would be expected from
simple capillarity arguments. We have explained both this shape and the amphiphile
concentrations in different regions of the system by considering the
balance of the surface tensions around the edge of the droplet.

Next, we studied the effect of the oil parameters on the droplet shape. We
found that, although reducing the incompatibility $\chi_\text{BO}$
causes the droplet to flatten and spread outwards into the bilayer, it
remains at least metastable even for very low $\chi_\text{BO}$. There
appears to be no clear threshold value of $\chi_\text{BO}$ for droplet
solutions to SCFT to exist. The droplets are also relatively
insensitive to changes in the oil molecule length. In fact, provided these molecules are longer than
the hydrophobic part of the amphiphile, their length has little effect
on the droplet shape. These observations provide some evidence that droplet
formation is a relatively robust effect and may therefore be a
reasonable explanation for phenomena such as the inclusion of
significant amounts of a second lipid species in lipid bilayers
\cite{khandelia}. Furthermore, the formation of droplets even in our simple
model indicates that this might be quite a general phenomenon and that
hydrophobic domains such as those seen in the molecular dynamics
simulations of Khandelia et al.\ \cite{khandelia} might be observed in a variety of systems.

The current work opens a number of interesting perspectives that could be
discussed within the
framework of self-consistent field theory. First, our
free energy calculations could be extended, to find the parameter
range where
the oil will form a droplet rather than spreading. Second, the
question of whether an optimum droplet size exists could be
addressed, perhaps by using a range of system sizes or by considering
the stability of a droplet with respect to two smaller droplets. We could also
study how likely the droplet is to split off from the bilayer, for example by comparing the free energies of
the bilayer-encapsulated droplet and a system of a droplet covered by
a monolayer in coexistence with a bilayer. Finally, droplets formed
from a second species of amphiphile could be studied, to bring our
calculations closer to the problem of lipid domains in bilayers
\cite{khandelia,hakumaki}. 
\section{Acknowledgements}
M.J.G. gratefully acknowledges funding from the EU under an FP7 Marie Curie fellowship.


\begin{thebibliography}{54}
\expandafter\ifx\csname natexlab\endcsname\relax\def\natexlab#1{#1}\fi
\expandafter\ifx\csname bibnamefont\endcsname\relax
  \def\bibnamefont#1{#1}\fi
\expandafter\ifx\csname bibfnamefont\endcsname\relax
  \def\bibfnamefont#1{#1}\fi
\expandafter\ifx\csname citenamefont\endcsname\relax
  \def\citenamefont#1{#1}\fi
\expandafter\ifx\csname url\endcsname\relax
  \def\url#1{\texttt{#1}}\fi
\expandafter\ifx\csname urlprefix\endcsname\relax\def\urlprefix{URL }\fi
\providecommand{\bibinfo}[2]{#2}
\providecommand{\eprint}[2][]{\url{#2}}

\bibitem[{\citenamefont{Zasadzinski et~al.}(2001)\citenamefont{Zasadzinski,
  Kisak, and Evans}}]{zasadzinski_rev}
\bibinfo{author}{\bibfnamefont{J.~A.} \bibnamefont{Zasadzinski}},
  \bibinfo{author}{\bibfnamefont{E.}~\bibnamefont{Kisak}}, \bibnamefont{and}
  \bibinfo{author}{\bibfnamefont{C.}~\bibnamefont{Evans}},
  \bibinfo{journal}{Curr.\ Opin.\ Colloid Interface Sci.}
  \textbf{\bibinfo{volume}{6}}, \bibinfo{pages}{85} (\bibinfo{year}{2001}).

\bibitem[{\citenamefont{Zidovska et~al.}(2009)\citenamefont{Zidovska, Ewert,
  Quispe, Carragher, Potter, and Safinya}}]{zidovska}
\bibinfo{author}{\bibfnamefont{A.}~\bibnamefont{Zidovska}},
  \bibinfo{author}{\bibfnamefont{K.~K.} \bibnamefont{Ewert}},
  \bibinfo{author}{\bibfnamefont{J.}~\bibnamefont{Quispe}},
  \bibinfo{author}{\bibfnamefont{B.}~\bibnamefont{Carragher}},
  \bibinfo{author}{\bibfnamefont{C.~S.} \bibnamefont{Potter}},
  \bibnamefont{and} \bibinfo{author}{\bibfnamefont{C.~R.}
  \bibnamefont{Safinya}}, \bibinfo{journal}{Langmuir}
  \textbf{\bibinfo{volume}{25}}, \bibinfo{pages}{2979} (\bibinfo{year}{2009}).

\bibitem[{\citenamefont{Sorre et~al.}(2009)\citenamefont{Sorre, Callan-Jones,
  Manneville, Nassoy, Joanny, Prost, Goud, and Bassereau}}]{sorre}
\bibinfo{author}{\bibfnamefont{B.}~\bibnamefont{Sorre}},
  \bibinfo{author}{\bibfnamefont{A.}~\bibnamefont{Callan-Jones}},
  \bibinfo{author}{\bibfnamefont{J.~B.} \bibnamefont{Manneville}},
  \bibinfo{author}{\bibfnamefont{P.}~\bibnamefont{Nassoy}},
  \bibinfo{author}{\bibfnamefont{J.~F.} \bibnamefont{Joanny}},
  \bibinfo{author}{\bibfnamefont{J.}~\bibnamefont{Prost}},
  \bibinfo{author}{\bibfnamefont{B.}~\bibnamefont{Goud}}, \bibnamefont{and}
  \bibinfo{author}{\bibfnamefont{P.}~\bibnamefont{Bassereau}},
  \bibinfo{journal}{Proc.\ Natl.\ Acad.\ Sci.\ U.S.A}
  \textbf{\bibinfo{volume}{106}}, \bibinfo{pages}{5622} (\bibinfo{year}{2009}).

\bibitem[{\citenamefont{Jain and Bates}(2003)}]{jain_bates}
\bibinfo{author}{\bibfnamefont{S.}~\bibnamefont{Jain}} \bibnamefont{and}
  \bibinfo{author}{\bibfnamefont{F.~S.} \bibnamefont{Bates}},
  \bibinfo{journal}{Science} \textbf{\bibinfo{volume}{300}},
  \bibinfo{pages}{460} (\bibinfo{year}{2003}).

\bibitem[{\citenamefont{Discher et~al.}(1999)\citenamefont{Discher, Won, Ege,
  Lee, Bates, Discher, and Hammer}}]{discher}
\bibinfo{author}{\bibfnamefont{B.~M.} \bibnamefont{Discher}},
  \bibinfo{author}{\bibfnamefont{Y.~Y.} \bibnamefont{Won}},
  \bibinfo{author}{\bibfnamefont{D.~S.} \bibnamefont{Ege}},
  \bibinfo{author}{\bibfnamefont{J.~C.~M.} \bibnamefont{Lee}},
  \bibinfo{author}{\bibfnamefont{F.~S.} \bibnamefont{Bates}},
  \bibinfo{author}{\bibfnamefont{D.~E.} \bibnamefont{Discher}},
  \bibnamefont{and} \bibinfo{author}{\bibfnamefont{D.~A.}
  \bibnamefont{Hammer}}, \bibinfo{journal}{Science}
  \textbf{\bibinfo{volume}{284}}, \bibinfo{pages}{1143} (\bibinfo{year}{1999}).

\bibitem[{\citenamefont{Discher and Eisenberg}(2002)}]{discher_eisenberg}
\bibinfo{author}{\bibfnamefont{D.~E.} \bibnamefont{Discher}} \bibnamefont{and}
  \bibinfo{author}{\bibfnamefont{A.}~\bibnamefont{Eisenberg}},
  \bibinfo{journal}{Science} \textbf{\bibinfo{volume}{297}},
  \bibinfo{pages}{967} (\bibinfo{year}{2002}).

\bibitem[{\citenamefont{Smart et~al.}(2010)\citenamefont{Smart, Ryan, Howse,
  and Battaglia}}]{smart}
\bibinfo{author}{\bibfnamefont{T.~P.} \bibnamefont{Smart}},
  \bibinfo{author}{\bibfnamefont{A.~J.} \bibnamefont{Ryan}},
  \bibinfo{author}{\bibfnamefont{J.~R.} \bibnamefont{Howse}}, \bibnamefont{and}
  \bibinfo{author}{\bibfnamefont{G.}~\bibnamefont{Battaglia}},
  \bibinfo{journal}{Langmuir} \textbf{\bibinfo{volume}{26}},
  \bibinfo{pages}{7425} (\bibinfo{year}{2010}).

\bibitem[{\citenamefont{Hamilton and Small}(1981)}]{hamilton}
\bibinfo{author}{\bibfnamefont{J.~A.} \bibnamefont{Hamilton}} \bibnamefont{and}
  \bibinfo{author}{\bibfnamefont{D.~M.} \bibnamefont{Small}},
  \bibinfo{journal}{Proc.\ Natl.\ Acad.\ Sci.\ U.S.A.}
  \textbf{\bibinfo{volume}{78}}, \bibinfo{pages}{6878} (\bibinfo{year}{1981}).

\bibitem[{\citenamefont{Hamilton}(1989)}]{hamilton2}
\bibinfo{author}{\bibfnamefont{J.~A.} \bibnamefont{Hamilton}},
  \bibinfo{journal}{Biochemistry} \textbf{\bibinfo{volume}{28}},
  \bibinfo{pages}{2514} (\bibinfo{year}{1989}).

\bibitem[{\citenamefont{Hamilton et~al.}(1991)\citenamefont{Hamilton, Fujito,
  and Hammer}}]{hamilton3}
\bibinfo{author}{\bibfnamefont{J.~A.} \bibnamefont{Hamilton}},
  \bibinfo{author}{\bibfnamefont{D.~T.} \bibnamefont{Fujito}},
  \bibnamefont{and} \bibinfo{author}{\bibfnamefont{C.~F.}
  \bibnamefont{Hammer}}, \bibinfo{journal}{Biochemistry}
  \textbf{\bibinfo{volume}{30}}, \bibinfo{pages}{2894} (\bibinfo{year}{1991}).

\bibitem[{\citenamefont{Khandelia et~al.}(2010)\citenamefont{Khandelia,
  Duelund, Pakkanen, and Ipsen}}]{khandelia}
\bibinfo{author}{\bibfnamefont{H.}~\bibnamefont{Khandelia}},
  \bibinfo{author}{\bibfnamefont{L.}~\bibnamefont{Duelund}},
  \bibinfo{author}{\bibfnamefont{K.~I.} \bibnamefont{Pakkanen}},
  \bibnamefont{and} \bibinfo{author}{\bibfnamefont{J.~H.} \bibnamefont{Ipsen}},
  \bibinfo{journal}{PLoS ONE} \textbf{\bibinfo{volume}{5}},
  \bibinfo{pages}{e12811} (\bibinfo{year}{2010}).

\bibitem[{\citenamefont{Ohsaki et~al.}(2009)\citenamefont{Ohsaki, Cheng,
  Suzuki, Shinohara, Fujita, and Fujimoto}}]{ohsaki}
\bibinfo{author}{\bibfnamefont{Y.}~\bibnamefont{Ohsaki}},
  \bibinfo{author}{\bibfnamefont{J.}~\bibnamefont{Cheng}},
  \bibinfo{author}{\bibfnamefont{M.}~\bibnamefont{Suzuki}},
  \bibinfo{author}{\bibfnamefont{Y.}~\bibnamefont{Shinohara}},
  \bibinfo{author}{\bibfnamefont{A.}~\bibnamefont{Fujita}}, \bibnamefont{and}
  \bibinfo{author}{\bibfnamefont{T.}~\bibnamefont{Fujimoto}},
  \bibinfo{journal}{Biochim.\ Biophys.\ Acta} \textbf{\bibinfo{volume}{1791}},
  \bibinfo{pages}{399} (\bibinfo{year}{2009}).

\bibitem[{\citenamefont{Zanghellini et~al.}(2010)\citenamefont{Zanghellini,
  Wodlei, and von Gr\"{u}nberg}}]{zanghellini}
\bibinfo{author}{\bibfnamefont{J.}~\bibnamefont{Zanghellini}},
  \bibinfo{author}{\bibfnamefont{F.}~\bibnamefont{Wodlei}}, \bibnamefont{and}
  \bibinfo{author}{\bibfnamefont{H.~H.} \bibnamefont{von Gr\"{u}nberg}},
  \bibinfo{journal}{J. Theor.\ Biol.} \textbf{\bibinfo{volume}{264}},
  \bibinfo{pages}{952} (\bibinfo{year}{2010}).

\bibitem[{\citenamefont{Ferretti et~al.}(1999)\citenamefont{Ferretti, Knijn,
  Iorio, Pulciani, Giambenedetti, Molinari, Meschini, Stringaro, Calcabrini,
  Freitas et~al.}}]{ferretti}
\bibinfo{author}{\bibfnamefont{A.}~\bibnamefont{Ferretti}},
  \bibinfo{author}{\bibfnamefont{A.}~\bibnamefont{Knijn}},
  \bibinfo{author}{\bibfnamefont{E.}~\bibnamefont{Iorio}},
  \bibinfo{author}{\bibfnamefont{S.}~\bibnamefont{Pulciani}},
  \bibinfo{author}{\bibfnamefont{M.}~\bibnamefont{Giambenedetti}},
  \bibinfo{author}{\bibfnamefont{A.}~\bibnamefont{Molinari}},
  \bibinfo{author}{\bibfnamefont{S.}~\bibnamefont{Meschini}},
  \bibinfo{author}{\bibfnamefont{A.}~\bibnamefont{Stringaro}},
  \bibinfo{author}{\bibfnamefont{A.}~\bibnamefont{Calcabrini}},
  \bibinfo{author}{\bibfnamefont{I.}~\bibnamefont{Freitas}},
  \bibnamefont{et~al.}, \bibinfo{journal}{Biochim.\ Biophys.\ Acta}
  \textbf{\bibinfo{volume}{1438}}, \bibinfo{pages}{329} (\bibinfo{year}{1999}).

\bibitem[{\citenamefont{May et~al.}(1986)\citenamefont{May, Wright, Holmes,
  Williams, Smith, Wright, fox, and Mountford}}]{may}
\bibinfo{author}{\bibfnamefont{G.~L.} \bibnamefont{May}},
  \bibinfo{author}{\bibfnamefont{L.~C.} \bibnamefont{Wright}},
  \bibinfo{author}{\bibfnamefont{K.~T.} \bibnamefont{Holmes}},
  \bibinfo{author}{\bibfnamefont{P.~G.} \bibnamefont{Williams}},
  \bibinfo{author}{\bibfnamefont{I.~C.~P.} \bibnamefont{Smith}},
  \bibinfo{author}{\bibfnamefont{P.~E.} \bibnamefont{Wright}},
  \bibinfo{author}{\bibfnamefont{R.~M.} \bibnamefont{fox}}, \bibnamefont{and}
  \bibinfo{author}{\bibfnamefont{C.~E.} \bibnamefont{Mountford}},
  \bibinfo{journal}{J. Biol.\ Chem.} \textbf{\bibinfo{volume}{261}},
  \bibinfo{pages}{3048} (\bibinfo{year}{1986}).

\bibitem[{\citenamefont{Hakum\"{a}ki and Kauppinen}(2000)}]{hakumaki}
\bibinfo{author}{\bibfnamefont{J.~M.} \bibnamefont{Hakum\"{a}ki}}
  \bibnamefont{and} \bibinfo{author}{\bibfnamefont{R.~A.}
  \bibnamefont{Kauppinen}}, \bibinfo{journal}{Trends in Biochemical Sciences}
  \textbf{\bibinfo{volume}{25}}, \bibinfo{pages}{357} (\bibinfo{year}{2000}).

\bibitem[{\citenamefont{Hayward et~al.}(2006)\citenamefont{Hayward, Utada, Dan,
  and Weitz}}]{hayward}
\bibinfo{author}{\bibfnamefont{R.~C.} \bibnamefont{Hayward}},
  \bibinfo{author}{\bibfnamefont{A.~S.} \bibnamefont{Utada}},
  \bibinfo{author}{\bibfnamefont{N.}~\bibnamefont{Dan}}, \bibnamefont{and}
  \bibinfo{author}{\bibfnamefont{D.~A.} \bibnamefont{Weitz}},
  \bibinfo{journal}{Langmuir} \textbf{\bibinfo{volume}{22}},
  \bibinfo{pages}{4457} (\bibinfo{year}{2006}).

\bibitem[{\citenamefont{Shum et~al.}(2008)\citenamefont{Shum, Lee, Yoon,
  Kodger, and Weitz}}]{shum}
\bibinfo{author}{\bibfnamefont{H.~C.} \bibnamefont{Shum}},
  \bibinfo{author}{\bibfnamefont{D.}~\bibnamefont{Lee}},
  \bibinfo{author}{\bibfnamefont{I.}~\bibnamefont{Yoon}},
  \bibinfo{author}{\bibfnamefont{T.}~\bibnamefont{Kodger}}, \bibnamefont{and}
  \bibinfo{author}{\bibfnamefont{D.~A.} \bibnamefont{Weitz}},
  \bibinfo{journal}{Langmuir} \textbf{\bibinfo{volume}{24}},
  \bibinfo{pages}{7651} (\bibinfo{year}{2008}).

\bibitem[{\citenamefont{Thiele et~al.}(2010)\citenamefont{Thiele, Abate, Shum,
  Bachtler, F\"{o}rster, and Weitz}}]{thiele}
\bibinfo{author}{\bibfnamefont{J.}~\bibnamefont{Thiele}},
  \bibinfo{author}{\bibfnamefont{A.~R.} \bibnamefont{Abate}},
  \bibinfo{author}{\bibfnamefont{H.~C.} \bibnamefont{Shum}},
  \bibinfo{author}{\bibfnamefont{S.}~\bibnamefont{Bachtler}},
  \bibinfo{author}{\bibfnamefont{S.}~\bibnamefont{F\"{o}rster}},
  \bibnamefont{and} \bibinfo{author}{\bibfnamefont{D.~A.} \bibnamefont{Weitz}},
  \bibinfo{journal}{Small} \textbf{\bibinfo{volume}{6}}, \bibinfo{pages}{1723}
  (\bibinfo{year}{2010}).

\bibitem[{\citenamefont{Onaca et~al.}(2009)\citenamefont{Onaca, Enea, Hughes,
  and Meier}}]{onaca}
\bibinfo{author}{\bibfnamefont{O.}~\bibnamefont{Onaca}},
  \bibinfo{author}{\bibfnamefont{R.}~\bibnamefont{Enea}},
  \bibinfo{author}{\bibfnamefont{D.~W.} \bibnamefont{Hughes}},
  \bibnamefont{and} \bibinfo{author}{\bibfnamefont{W.}~\bibnamefont{Meier}},
  \bibinfo{journal}{Macromol.\ Biosci.} \textbf{\bibinfo{volume}{9}},
  \bibinfo{pages}{129} (\bibinfo{year}{2009}).

\bibitem[{\citenamefont{Meng et~al.}(2009)\citenamefont{Meng, Zhong, and
  Feijen}}]{meng}
\bibinfo{author}{\bibfnamefont{F.}~\bibnamefont{Meng}},
  \bibinfo{author}{\bibfnamefont{Z.}~\bibnamefont{Zhong}}, \bibnamefont{and}
  \bibinfo{author}{\bibfnamefont{J.}~\bibnamefont{Feijen}},
  \bibinfo{journal}{Biomacromolecules} \textbf{\bibinfo{volume}{10}},
  \bibinfo{pages}{197} (\bibinfo{year}{2009}).

\bibitem[{\citenamefont{Li et~al.}(2010)\citenamefont{Li, Li, Wan, and
  Hou}}]{li}
\bibinfo{author}{\bibfnamefont{D.}~\bibnamefont{Li}},
  \bibinfo{author}{\bibfnamefont{C.}~\bibnamefont{Li}},
  \bibinfo{author}{\bibfnamefont{G.}~\bibnamefont{Wan}}, \bibnamefont{and}
  \bibinfo{author}{\bibfnamefont{W.}~\bibnamefont{Hou}},
  \bibinfo{journal}{Colloids Surf.\ A} \textbf{\bibinfo{volume}{372}},
  \bibinfo{pages}{1} (\bibinfo{year}{2010}).

\bibitem[{\citenamefont{Mueller et~al.}(2009)\citenamefont{Mueller, Koynov,
  Fischer, Hartmann, Pierrat, Basch\'{e}, and Maskos}}]{mueller}
\bibinfo{author}{\bibfnamefont{W.}~\bibnamefont{Mueller}},
  \bibinfo{author}{\bibfnamefont{K.}~\bibnamefont{Koynov}},
  \bibinfo{author}{\bibfnamefont{K.}~\bibnamefont{Fischer}},
  \bibinfo{author}{\bibfnamefont{S.}~\bibnamefont{Hartmann}},
  \bibinfo{author}{\bibfnamefont{S.}~\bibnamefont{Pierrat}},
  \bibinfo{author}{\bibfnamefont{T.}~\bibnamefont{Basch\'{e}}},
  \bibnamefont{and} \bibinfo{author}{\bibfnamefont{M.}~\bibnamefont{Maskos}},
  \bibinfo{journal}{Macromolecules} \textbf{\bibinfo{volume}{42}},
  \bibinfo{pages}{357} (\bibinfo{year}{2009}).

\bibitem[{\citenamefont{Qin et~al.}(2006)\citenamefont{Qin, Geng, Discher, and
  Yang}}]{qin}
\bibinfo{author}{\bibfnamefont{S.}~\bibnamefont{Qin}},
  \bibinfo{author}{\bibfnamefont{Y.}~\bibnamefont{Geng}},
  \bibinfo{author}{\bibfnamefont{D.~E.} \bibnamefont{Discher}},
  \bibnamefont{and} \bibinfo{author}{\bibfnamefont{S.}~\bibnamefont{Yang}},
  \bibinfo{journal}{Adv.\ Mater.} \textbf{\bibinfo{volume}{18}},
  \bibinfo{pages}{2905} (\bibinfo{year}{2006}).

\bibitem[{\citenamefont{Zasadzinski et~al.}(2011)\citenamefont{Zasadzinski,
  Wong, Forbes, Braun, and Wu}}]{zasadzinski}
\bibinfo{author}{\bibfnamefont{J.~A.} \bibnamefont{Zasadzinski}},
  \bibinfo{author}{\bibfnamefont{B.}~\bibnamefont{Wong}},
  \bibinfo{author}{\bibfnamefont{N.}~\bibnamefont{Forbes}},
  \bibinfo{author}{\bibfnamefont{G.}~\bibnamefont{Braun}}, \bibnamefont{and}
  \bibinfo{author}{\bibfnamefont{G.}~\bibnamefont{Wu}},
  \bibinfo{journal}{Curr.\ Opin.\ Colloid Interface Sci.}
  \textbf{\bibinfo{volume}{16}}, \bibinfo{pages}{203} (\bibinfo{year}{2011}).

\bibitem[{\citenamefont{Chen et~al.}(2010)\citenamefont{Chen, Meng, Cheng, and
  Zhong}}]{chen}
\bibinfo{author}{\bibfnamefont{W.}~\bibnamefont{Chen}},
  \bibinfo{author}{\bibfnamefont{F.}~\bibnamefont{Meng}},
  \bibinfo{author}{\bibfnamefont{R.}~\bibnamefont{Cheng}}, \bibnamefont{and}
  \bibinfo{author}{\bibfnamefont{Z.}~\bibnamefont{Zhong}}, \bibinfo{journal}{J.
  Controlled Release} \textbf{\bibinfo{volume}{142}}, \bibinfo{pages}{40}
  (\bibinfo{year}{2010}).

\bibitem[{\citenamefont{Edwards}(1965)}]{edwards}
\bibinfo{author}{\bibfnamefont{S.~F.} \bibnamefont{Edwards}},
  \bibinfo{journal}{Proc.\ Phys.\ Soc.} \textbf{\bibinfo{volume}{85}},
  \bibinfo{pages}{613} (\bibinfo{year}{1965}).

\bibitem[{\citenamefont{Maniadis et~al.}(2007)\citenamefont{Maniadis, Lookman,
  Kober, and Rasmussen}}]{maniadis}
\bibinfo{author}{\bibfnamefont{P.}~\bibnamefont{Maniadis}},
  \bibinfo{author}{\bibfnamefont{T.}~\bibnamefont{Lookman}},
  \bibinfo{author}{\bibfnamefont{E.~M.} \bibnamefont{Kober}}, \bibnamefont{and}
  \bibinfo{author}{\bibfnamefont{K.~O.} \bibnamefont{Rasmussen}},
  \bibinfo{journal}{Phys.\ Rev.\ Lett.} \textbf{\bibinfo{volume}{99}},
  \bibinfo{pages}{048302} (\bibinfo{year}{2007}).

\bibitem[{\citenamefont{Drolet and Fredrickson}(1999)}]{drolet_fredrickson}
\bibinfo{author}{\bibfnamefont{F.}~\bibnamefont{Drolet}} \bibnamefont{and}
  \bibinfo{author}{\bibfnamefont{G.~H.} \bibnamefont{Fredrickson}},
  \bibinfo{journal}{Phys.\ Rev.\ Lett.} \textbf{\bibinfo{volume}{83}},
  \bibinfo{pages}{4317} (\bibinfo{year}{1999}).

\bibitem[{\citenamefont{Matsen}(2006)}]{matsen_book}
\bibinfo{author}{\bibfnamefont{M.~W.} \bibnamefont{Matsen}},
  \emph{\bibinfo{title}{Soft Matter}} (\bibinfo{publisher}{Wiley-VCH},
  \bibinfo{address}{Weinheim}, \bibinfo{year}{2006}),
  chap.~\bibinfo{chapter}{2}.

\bibitem[{\citenamefont{Duque}(2003)}]{duque}
\bibinfo{author}{\bibfnamefont{D.}~\bibnamefont{Duque}}, \bibinfo{journal}{J.
  Chem.\ Phys.} \textbf{\bibinfo{volume}{119}}, \bibinfo{pages}{5701}
  (\bibinfo{year}{2003}).

\bibitem[{\citenamefont{Katsov et~al.}(2004)\citenamefont{Katsov, M\"{u}ller,
  and Schick}}]{katsov1}
\bibinfo{author}{\bibfnamefont{K.}~\bibnamefont{Katsov}},
  \bibinfo{author}{\bibfnamefont{M.}~\bibnamefont{M\"{u}ller}},
  \bibnamefont{and} \bibinfo{author}{\bibfnamefont{M.}~\bibnamefont{Schick}},
  \bibinfo{journal}{Biophys.\ J.} \textbf{\bibinfo{volume}{87}},
  \bibinfo{pages}{3277} (\bibinfo{year}{2004}).

\bibitem[{\citenamefont{Cavallo et~al.}(2006)\citenamefont{Cavallo, M\"{u}ller,
  and Binder}}]{cavallo}
\bibinfo{author}{\bibfnamefont{A.}~\bibnamefont{Cavallo}},
  \bibinfo{author}{\bibfnamefont{M.}~\bibnamefont{M\"{u}ller}},
  \bibnamefont{and} \bibinfo{author}{\bibfnamefont{K.}~\bibnamefont{Binder}},
  \bibinfo{journal}{Macromolecules} \textbf{\bibinfo{volume}{39}},
  \bibinfo{pages}{9539} (\bibinfo{year}{2006}).

\bibitem[{\citenamefont{Schuetz et~al.}(2011)\citenamefont{Schuetz, Greenall,
  Bent, Furzeland, Atkins, Butler, McLeish, and Buzza}}]{schuetz}
\bibinfo{author}{\bibfnamefont{P.}~\bibnamefont{Schuetz}},
  \bibinfo{author}{\bibfnamefont{M.~J.} \bibnamefont{Greenall}},
  \bibinfo{author}{\bibfnamefont{J.}~\bibnamefont{Bent}},
  \bibinfo{author}{\bibfnamefont{S.}~\bibnamefont{Furzeland}},
  \bibinfo{author}{\bibfnamefont{D.}~\bibnamefont{Atkins}},
  \bibinfo{author}{\bibfnamefont{M.~F.} \bibnamefont{Butler}},
  \bibinfo{author}{\bibfnamefont{T.~C.~B.} \bibnamefont{McLeish}},
  \bibnamefont{and} \bibinfo{author}{\bibfnamefont{D.~M.~A.}
  \bibnamefont{Buzza}}, \bibinfo{journal}{Soft Matter}
  \textbf{\bibinfo{volume}{7}}, \bibinfo{pages}{749} (\bibinfo{year}{2011}).

\bibitem[{\citenamefont{Werner et~al.}(1999)\citenamefont{Werner, M\"uller,
  Schmid, and Binder}}]{werner}
\bibinfo{author}{\bibfnamefont{A.}~\bibnamefont{Werner}},
  \bibinfo{author}{\bibfnamefont{M.}~\bibnamefont{M\"uller}},
  \bibinfo{author}{\bibfnamefont{F.}~\bibnamefont{Schmid}}, \bibnamefont{and}
  \bibinfo{author}{\bibfnamefont{K.}~\bibnamefont{Binder}},
  \bibinfo{journal}{J. Chem.\ Phys.} \textbf{\bibinfo{volume}{110}},
  \bibinfo{pages}{1221} (\bibinfo{year}{1999}).

\bibitem[{\citenamefont{M\"{u}ller and Gompper}(2002)}]{mueller_gompper}
\bibinfo{author}{\bibfnamefont{M.}~\bibnamefont{M\"{u}ller}} \bibnamefont{and}
  \bibinfo{author}{\bibfnamefont{G.}~\bibnamefont{Gompper}},
  \bibinfo{journal}{Phys.\ Rev.\ E} \textbf{\bibinfo{volume}{66}},
  \bibinfo{pages}{041805} (\bibinfo{year}{2002}).

\bibitem[{\citenamefont{Wang et~al.}(2010)\citenamefont{Wang, Guo, An,
  M\"uller, and Wang}}]{wang}
\bibinfo{author}{\bibfnamefont{J.~F.} \bibnamefont{Wang}},
  \bibinfo{author}{\bibfnamefont{K.~K.} \bibnamefont{Guo}},
  \bibinfo{author}{\bibfnamefont{L.~J.} \bibnamefont{An}},
  \bibinfo{author}{\bibfnamefont{M.}~\bibnamefont{M\"uller}}, \bibnamefont{and}
  \bibinfo{author}{\bibfnamefont{Z.~G.} \bibnamefont{Wang}},
  \bibinfo{journal}{Macromolecules} \textbf{\bibinfo{volume}{43}},
  \bibinfo{pages}{2037} (\bibinfo{year}{2010}).

\bibitem[{\citenamefont{Denesyuk and Gompper}(2006)}]{denesyuk}
\bibinfo{author}{\bibfnamefont{N.~A.} \bibnamefont{Denesyuk}} \bibnamefont{and}
  \bibinfo{author}{\bibfnamefont{G.}~\bibnamefont{Gompper}},
  \bibinfo{journal}{Macromolecules} \textbf{\bibinfo{volume}{39}},
  \bibinfo{pages}{5497} (\bibinfo{year}{2006}).

\bibitem[{\citenamefont{Wijmans and Linse}(1995)}]{wijmans_linse}
\bibinfo{author}{\bibfnamefont{C.~M.} \bibnamefont{Wijmans}} \bibnamefont{and}
  \bibinfo{author}{\bibfnamefont{P.}~\bibnamefont{Linse}},
  \bibinfo{journal}{Langmuir} \textbf{\bibinfo{volume}{11}},
  \bibinfo{pages}{3748} (\bibinfo{year}{1995}).

\bibitem[{\citenamefont{Leermakers and
  Scheutjens}(1990)}]{leermakers_scheutjens-shape}
\bibinfo{author}{\bibfnamefont{F.~A.~M.} \bibnamefont{Leermakers}}
  \bibnamefont{and} \bibinfo{author}{\bibfnamefont{J.~M. H.~M.}
  \bibnamefont{Scheutjens}}, \bibinfo{journal}{J. Colloid Interface Sci.}
  \textbf{\bibinfo{volume}{136}}, \bibinfo{pages}{231} (\bibinfo{year}{1990}).

\bibitem[{\citenamefont{Fredrickson}(2006)}]{fredrickson_book}
\bibinfo{author}{\bibfnamefont{G.~H.} \bibnamefont{Fredrickson}},
  \emph{\bibinfo{title}{The Equilibrium Theory of Inhomogeneous Polymers}}
  (\bibinfo{publisher}{Oxford University Press}, \bibinfo{address}{Oxford},
  \bibinfo{year}{2006}).

\bibitem[{\citenamefont{Schmid}(1998)}]{schmid_scf_rev}
\bibinfo{author}{\bibfnamefont{F.}~\bibnamefont{Schmid}}, \bibinfo{journal}{J.
  Phys.: Condens.\ Matter} \textbf{\bibinfo{volume}{10}}, \bibinfo{pages}{8105}
  (\bibinfo{year}{1998}).

\bibitem[{\citenamefont{Greenall and Gompper}(2011)}]{gg}
\bibinfo{author}{\bibfnamefont{M.~J.} \bibnamefont{Greenall}} \bibnamefont{and}
  \bibinfo{author}{\bibfnamefont{G.}~\bibnamefont{Gompper}},
  \bibinfo{journal}{Langmuir} \textbf{\bibinfo{volume}{27}},
  \bibinfo{pages}{3416} (\bibinfo{year}{2011}).

\bibitem[{\citenamefont{Jones}(2002)}]{jones_book}
\bibinfo{author}{\bibfnamefont{R.~A.~L.} \bibnamefont{Jones}},
  \emph{\bibinfo{title}{Soft Condensed Matter}} (\bibinfo{publisher}{Oxford
  University Press}, \bibinfo{address}{Oxford}, \bibinfo{year}{2002}).

\bibitem[{\citenamefont{Greenall et~al.}(2011)\citenamefont{Greenall, Schuetz,
  Furzeland, Atkins, Buzza, Butler, and McLeish}}]{schuetz2}
\bibinfo{author}{\bibfnamefont{M.~J.} \bibnamefont{Greenall}},
  \bibinfo{author}{\bibfnamefont{P.}~\bibnamefont{Schuetz}},
  \bibinfo{author}{\bibfnamefont{S.}~\bibnamefont{Furzeland}},
  \bibinfo{author}{\bibfnamefont{D.}~\bibnamefont{Atkins}},
  \bibinfo{author}{\bibfnamefont{D.~M.~A.} \bibnamefont{Buzza}},
  \bibinfo{author}{\bibfnamefont{M.~F.} \bibnamefont{Butler}},
  \bibnamefont{and} \bibinfo{author}{\bibfnamefont{T.~C.~B.}
  \bibnamefont{McLeish}}, \bibinfo{journal}{Macromolecules}
  \textbf{\bibinfo{volume}{44}}, \bibinfo{pages}{5510} (\bibinfo{year}{2011}).

\bibitem[{\citenamefont{Greenall
  et~al.}(2009{\natexlab{a}})\citenamefont{Greenall, Buzza, and
  McLeish}}]{gbm_jcp}
\bibinfo{author}{\bibfnamefont{M.~J.} \bibnamefont{Greenall}},
  \bibinfo{author}{\bibfnamefont{D.~M.~A.} \bibnamefont{Buzza}},
  \bibnamefont{and} \bibinfo{author}{\bibfnamefont{T.~C.~B.}
  \bibnamefont{McLeish}}, \bibinfo{journal}{J. Chem.\ Phys.}
  \textbf{\bibinfo{volume}{131}}, \bibinfo{pages}{034904}
  (\bibinfo{year}{2009}{\natexlab{a}}).

\bibitem[{\citenamefont{Press et~al.}(1992)\citenamefont{Press, Flannery,
  Teukolsky, and Vetterling}}]{num_rec}
\bibinfo{author}{\bibfnamefont{W.~H.} \bibnamefont{Press}},
  \bibinfo{author}{\bibfnamefont{B.~P.} \bibnamefont{Flannery}},
  \bibinfo{author}{\bibfnamefont{S.~A.} \bibnamefont{Teukolsky}},
  \bibnamefont{and} \bibinfo{author}{\bibfnamefont{W.~T.}
  \bibnamefont{Vetterling}}, \emph{\bibinfo{title}{Numerical Recipes in C}}
  (\bibinfo{publisher}{Cambridge University Press},
  \bibinfo{address}{Cambridge}, \bibinfo{year}{1992}), \bibinfo{edition}{2nd}
  ed.

\bibitem[{\citenamefont{Matsen}(2004)}]{matsen2004}
\bibinfo{author}{\bibfnamefont{M.~W.} \bibnamefont{Matsen}},
  \bibinfo{journal}{J. Chem.\ Phys.} \textbf{\bibinfo{volume}{121}},
  \bibinfo{pages}{1938} (\bibinfo{year}{2004}).

\bibitem[{\citenamefont{Greenall
  et~al.}(2009{\natexlab{b}})\citenamefont{Greenall, Buzza, and
  McLeish}}]{gbm_macro}
\bibinfo{author}{\bibfnamefont{M.~J.} \bibnamefont{Greenall}},
  \bibinfo{author}{\bibfnamefont{D.~M.~A.} \bibnamefont{Buzza}},
  \bibnamefont{and} \bibinfo{author}{\bibfnamefont{T.~C.~B.}
  \bibnamefont{McLeish}}, \bibinfo{journal}{Macromolecules}
  \textbf{\bibinfo{volume}{42}}, \bibinfo{pages}{5873}
  (\bibinfo{year}{2009}{\natexlab{b}}).

\bibitem[{\citenamefont{Schmid}(2011)}]{boudenne_book}
\bibinfo{author}{\bibfnamefont{F.}~\bibnamefont{Schmid}}, in
  \emph{\bibinfo{booktitle}{Handbook of Multiphase Polymer Systems}}, edited by
  \bibinfo{editor}{\bibfnamefont{A.}~\bibnamefont{Boudenne}}
  (\bibinfo{publisher}{John Wiley and Sons}, \bibinfo{address}{Chichester},
  \bibinfo{year}{2011}), chap.~\bibinfo{chapter}{3}.

\bibitem[{\citenamefont{L\'{e}ger and Joanny}(1992)}]{leger}
\bibinfo{author}{\bibfnamefont{L.}~\bibnamefont{L\'{e}ger}} \bibnamefont{and}
  \bibinfo{author}{\bibfnamefont{J.~F.} \bibnamefont{Joanny}},
  \bibinfo{journal}{Rep.\ Prog.\ Phys.} \textbf{\bibinfo{volume}{55}},
  \bibinfo{pages}{431} (\bibinfo{year}{1992}).

\bibitem[{\citenamefont{Rowlinson and Widom}(1982)}]{rowlinson_widom}
\bibinfo{author}{\bibfnamefont{J.~S.} \bibnamefont{Rowlinson}}
  \bibnamefont{and} \bibinfo{author}{\bibfnamefont{B.}~\bibnamefont{Widom}},
  \emph{\bibinfo{title}{Molecular Theory of Capillarity}}
  (\bibinfo{publisher}{Clarendon Press}, \bibinfo{address}{Oxford},
  \bibinfo{year}{1982}).

\bibitem[{\citenamefont{Landau and Lifshitz}(1987)}]{landau_lifshitz}
\bibinfo{author}{\bibfnamefont{L.~D.} \bibnamefont{Landau}} \bibnamefont{and}
  \bibinfo{author}{\bibfnamefont{E.~M.} \bibnamefont{Lifshitz}},
  \emph{\bibinfo{title}{Fluid Mechanics}} (\bibinfo{publisher}{Pergamon Press},
  \bibinfo{address}{Oxford}, \bibinfo{year}{1987}), \bibinfo{edition}{2nd} ed.

\bibitem[{\citenamefont{Kusumaatmaja and Lipowsky}(2011)}]{kusumaatmaja}
\bibinfo{author}{\bibfnamefont{H.}~\bibnamefont{Kusumaatmaja}}
  \bibnamefont{and} \bibinfo{author}{\bibfnamefont{R.}~\bibnamefont{Lipowsky}},
  \bibinfo{journal}{Soft Matter} \textbf{\bibinfo{volume}{7}},
  \bibinfo{pages}{6914} (\bibinfo{year}{2011}).

\end{thebibliography}
\end{document}